\documentclass{article}
\usepackage{spconf,amsmath,graphicx}
\usepackage{booktabs}
\usepackage{adjustbox}
\usepackage{multirow}
\usepackage{textcomp}
\usepackage{caption}
\usepackage{subcaption}

\title{End-to-End Language Identification using Multi-Head Self-Attention and 1D Convolutional Neural Networks}
\name{Krishna D N, Ankita Patil}
\address{
  FreshWorks Inc.,
  HashCut Inc.
  }

\begin{document}
%\ninept
%
\maketitle
\begin{abstract}
In this work, we propose a new approach for language identification using multi-head self-attention combined with raw waveform based 1D convolutional neural networks for Indian languages. Our approach uses an encoder, multi-head self-attention, and a statistics pooling layer. The encoder learns features directly from raw waveforms using 1D convolution kernels and an LSTM layer. The LSTM layer captures temporal information between the features extracted by the 1D convolutional layer. The multi-head self-attention layer takes outputs of the LSTM layer and applies self-attention mechanisms on these features with M different heads. This process helps the model give more weightage to the more useful features and less weightage to the less relevant features. Finally, the frame-level features are combined using a statistics pooling layer to extract the utterance-level feature vector label prediction. We conduct all our experiments on the ~373 hrs of audio data for eight different Indian languages. Our experiments show that our approach outperforms the baseline model by an absolute 3.69\% improvement in F1-score and achieves the best F1-score of 95.90\%. Our approach also shows that using raw waveform models gets a 1.7\% improvement in performance compared to the models built using handcrafted features. 
\end{abstract}
\section{INTRODUCTION}
\label{sec:intro}
A recent development in the area of deep neural networks has shown tremendous improvements in speech systems, including speech recognition[1,2,3], emotion recognition[20], speaker identification [13]. Previously, the language identification field was dominated by i-vectors [5], which uses Gaussian mixture models traditionally. Even today, i-vectors are considered to be the best model in the case of low data regime. However, recent developments in the field of deep learning show that deep neural networks are one of the dominating approaches in language identification. Previously, W. Geng et al. [4] proposed to use deep features extracted from a neural network trained for speech recognition and showed that deep neural models are capable of obtaining excellent performance over the classical systems[7,9,10,15,19]. Recently, Time-delay neural networks(TDNN) have shown excellent performance for speech recognition tasks [21]. X-vector [13] built for speaker identification was used for language identification [14], and its shown to be one of the best methods for language identification. The recent trend in sequence to sequence mapping problems [18] involves the attention mechanism [16]. The attention mechanism is one of the very well known techniques being used in sequence mapping problems [17], and today's state of the art speech recognition models are built using the attention-based models. These models process sequential inputs by iteratively selecting relevant features using attention technique. Due to the efficiency of the attention technique for sequence mapping problems, A. Vaswani et al. [22] propose to use stacks of attention layers alone. They have shown remarkable results in machine translation.
Motivated by the work [22], this paper proposes to use multi-head self-attention in combination with 1D convolutional neural network front-end processing for language identification. Ours proposed model takes raw waveform as input directly and extracts features that are useful for the LID task. The model consists of a sequence of Residual blocks [23] of 1D convolutional layers to extract features from the raw audio. Since the convolutional layer does not have the capability to capture temporal information, we use an LSTM layer on top of the convolutional network to capture time-varying information from the input. Finally, the LSTM output feature sequence is fed into a multi-head self-attention block consisting of multiple attention heads to select important features from different parts of the input feature sequence using attention weighting. Finally, an utterance level feature vector is generating using a statistics pooling layer, and we classify the utterance level feature to predict the class label.The organization of the paper is as follows. In section 2, we explain our proposed approach in detail. In section 3, we give a detailed analysis of the dataset collection and curation process, and in section 4, we explain our experimental setup in detail. Finally, in section 5, we describe our results.
\section{Proposed method}
In this section, we explain our proposed approach in detail. The detailed model architecture is shown in Figure 1. Our model consists of 3 main stages, 1) An Encoder layer, which includes multiple 1D convolutional layers with residual connections, an LSTM layer, 2) Multi-head self-attention layer to select important features for language identification using attention weighting, and 3)statistic pooling layer to obtain utterance level feature vector for classification. The model takes raw audio waveform as input and applies initial 1D convolution operations along with 1D max-pooling, as shown in Figure 1. The initial convolutional layer features will go through a series of there 1D Residual blocks followed by an LSTM layer. We then use multi-head self-attention to extract relevant features from different parts of the input. The statistics pooling layer generates an utterance level feature vector containing language discriminative properties. The output of the statistics pooling layer gives us a single feature vector called the utterance level feature vector. This utterance level feature vector is fed into a projection layer followed by a softmax layer to predict the class label. We explain the details of each of these blocks in the following section.

\begin{figure}[!htbp]
  \centering
 \minipage{0.4\textwidth}
  \includegraphics[width=\linewidth]{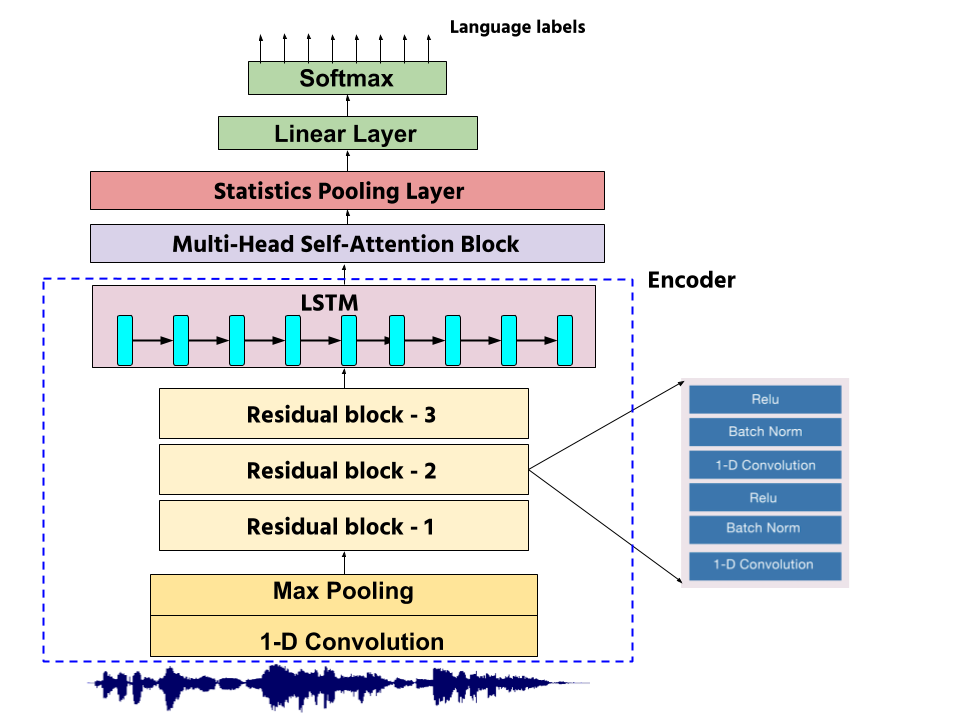}
  \caption{Proposed model architecture}
    \label{fig:tree-net}
  \endminipage\hfill
\end{figure}

\subsection{Encoder}
The encoder of our model consists of a series of three residual blocks combined with a single LSTM layer, as shown in Figure 1. The encoder takes a raw waveform signal and applies an initial 1D convolution operation consisting of 64 filters of 1x7 filter size followed by a max-pooling operation. The max-pooling is applied with a kernel size of 1x3 with stride 2. After the initial convolution and max-pooling operation, we send the output a sequence of residual blocks. The details of a single residual block is shown in Figure 1. Each residual blocks operate with 1D convolution kernels of size 1x3. \textit{Residual Block -1} consist of 2*64 convolution kernels of size 1x3, \textit{Residual Block -2} consist of 2*128 convolution kernels of size 1x3 and \textit{Residual Block -3} contains 2*256 convolution kernels of size 1x3. The output of \textit{Residual Block -2} and \textit{Residual Block -3} will go through a 1D max-pooling operation. Since LSTMs are known to capture long-range dependencies between frames in speech, we use LSTM at the end of the final residual block to capture temporal information. The output of the final residual block \textit{Residual Block -3} after the max-pooling operation is sent to a single unidirectional LSTM layer with a hidden size of 256. Let $\boldsymbol{X_n=[x_1,x_2..x_n,...x_N]}$ be raw audio sequence with N samples.  

\begin{equation}
\boldsymbol{H^A} = \textsf{Encoder}(\boldsymbol{X_n})
\end{equation}

Where, \textsf{Encoder} is a mapping function which consists of initial 1D convolutional layer and max-pooling operation, sequence of there residual blocks \textit{Residual Block-1},\textit{Residual Block-2} and \textit{Residual Block-3} along with LSTM layer . After this operation, we obtain a feature sequence  $\boldsymbol{H^A=[h_1,h_2.....h_T]}$ of length T (T$<<$N). Typically after the convolution operation, $\boldsymbol{H^A}$ can be looked at as a feature matrix whose x-axis is a time dimension, and the y-axis is a feature dimension. The feature dimension in our case is 256, as the hidden layer size of the LSTM is the same.

\subsection{Multi-Head Self-Attention}
In this section, we describe multi-head self-attention in detail. It consists of 3 different linear blocks, one for query, one for key, and another for value. Each linear block consists of M independent linear layers. Where M is the number of heads. The multi-head attention block takes features $\boldsymbol{H^A=[h_1,h_2.....h_T]}$  from LSTM and applies linear transformation to create $\boldsymbol{Q_i}$, $\boldsymbol{K_i}$ and $\boldsymbol{V_i}$ using $\boldsymbol{i^{\text{th}}}$ linear layers where, $\boldsymbol{i=[1,2.....M]}$  and M is the total number of attention heads. The $\boldsymbol{Q_i}$, $\boldsymbol{K_i}$ and $\boldsymbol{V_i}$ are fed into scaled dot product attention layer. The scaled dot product attention $\boldsymbol{A_i}$ for $\boldsymbol{i^{\text{th}}}$ head is defined as follows.

\begin{equation}
\boldsymbol{A_i} = \textsf{Softmax}(\boldsymbol{\frac{Q_iK_i}{d_q}})\boldsymbol{V_i}
\end{equation}

Where $\boldsymbol{d_q}$ is the dimension of the query vector. We combine the attention output from all the heads using simple concatenation and feed into the feed-forward layer.

\begin{equation}
\boldsymbol{A} = \textsf{Concat}(\boldsymbol{A_1,A_2,A_3...A_i.....A_M}){W_0}
\end{equation}

Where, $\boldsymbol{A_i}$ is a $\boldsymbol{d_q}$x$\boldsymbol{T}$ dimensional matrix.
Since the \textsf{Concat} operation is applied to the feature dimension of all the matrices, the final output attention matrix $\boldsymbol{A}$ from the multi-head attention block will have  $\boldsymbol{Md_q}$x$\boldsymbol{T}$ matrix dimensions.

The multi-head attention layer helps in finding features that are more relevant for language identification. The scaled dot product attention achieves this by giving more weighting to the more relevant features and less weighting to less relevant features. Due to the presence of multiple heads in the attention layer, this process selects features from different parts of the input and helps in obtaining better language classification performance.

\subsection{Statistics pooling}
The idea of the statistics pooling layer is similar to max pooling. In the case of statistics pooling, we compute the mean and standard deviation from frame-level features. The mean and standard deviation features are concatenated to create the utterance level feature vector, as described in the equation below. Let $\boldsymbol{A=[a_1,a_2.....a_T]}$  is the output from multi-head attention block.

\begin{equation}
\boldsymbol{P} = \textsf{Concat}({\textsf{mean}(\boldsymbol{A})},{\textsf{std}(\boldsymbol{A})})
\end{equation}

Where, $\boldsymbol{a_i}$ is a feature vector of dimension $\boldsymbol{M*d_q}$ and $\boldsymbol{P}$ is final pooled feature vector using statistics pooling layer. Since the dimension of the utterance level feature vector is $\boldsymbol{P}$ become bigger when M is large, we add a projection layer on top to the statistics pooling layer (Figure 1) in order to reduce the dimension of $\boldsymbol{P}$. We take the output from this projection layer to visualize the utterance level embeddings for different languages.

\section{Dataset}
In this section, we describe our data collection process. We collect and curate videos from Youtube\footnote{www.youtube.com} using manual labeling. We ask annotators to look for videos for eight languages in Youtube and manually verify it make sure the video does not contain multiple Indian languages. Most of these videos contain background noise or music signals. Sometime the video may contain a mix of English and other Indian languages due to code-mixing. We use an in-house speech v/s non-speech detection model to detect only the speech segments. We clip only the speech segments from every video and discard the non-speech part of the video. After preprocessing, our total dataset contains 373.27 hrs of audio data for 8 Indian languages. Our dataset includes \textit{Hindi}, \textit{English}, \textit{Kannada},\textit{Tamil}, \textit{Telugu}, \textit{Malayalam}, \textit{Gujarati} and \textit{Marathi}. These languages are officially spoken in the North and South regions of India. We split the dataset into training and evaluation part, and the statistics of the training and evaluation parts are shown in Table 1.

\begin{table}
\centering
  \label{tab:tasks}
  
  \begin{adjustbox}{max width=\textwidth}
  \begin{tabular}{|l|l|l|l|l|}
    \hline
    \multirow{2}{*}{Datset} &
     \multicolumn{2}{c}{Train} &
     \multicolumn{2}{c}{Eval} \\
    & Duration & Files & Duration & Files\\
    \hline
    English & 44.21 & 15963 & 11.28 & 4074 \\
    \hline
    Kannada & 35.95 & 12988 & 8.90 & 3216 \\
    \hline
    Gujarati & 30.24 & 10933 & 7.312 & 2642 \\
    \hline
    Hindi & 38.79 & 14004 & 9.72 & 3510 \\
    \hline
    Malayalam & 34.98 & 12636 & 8.94 & 3228 \\
    \hline
    Tamil & 63.15 & 22774 & 15.95 & 5753 \\
    \hline
    Telugu & 35.07 & 12666 & 8.54 & 3087 \\
    \hline
    Marathi & 16.23 & 5873 & 4.013 & 1449 \\
    \hline
  \end{tabular}
  \end{adjustbox}
  \caption{Train and evaluation splits for different languages (Duration is in Hrs)}
\end{table}

\section{Experiments}
We conduct all our experiments on in house dataset collected for 8 Indian languages. Our proposed model consists of an encoder, multi-head self-attention block, and statistics pooling layer followed by projection and a softmax layer. We randomly select a 4sec audio signal from each audio file during training. Since our data has a sampling rate of 16KHz, we get 64000 samples from every file during training. We feed a 1x64000 dimensional signal into our encoder. We conduct multiple experiments to see the effectiveness of the multi-head self-attention module for language identification. We first train a standalone 1D convolutional neural network model as the first baseline model. We refer this system as $\textit{ResNet}$. We also train a 1D convolutional neural network in combination with a unidirectional LSTM as a second baseline, and we refer to it as $\textit{ResNet-LSTM}$. Finally, our proposed model is built using a 1D convolutional neural network, LSTM, and multi-head self-attention. We refer to it as $\textit{ResNet-LSTM-MHA-Raw}$.

We conduct multiple experiments to see the effectiveness of our model on the duration of the audio during training. We train 3 different models $\textit{ResNet-LSTM-MHA-2Sec}$, $\textit{ResNet-LSTM-MHA-3Sec}$, $\textit{ResNet-LSTM-MHA-4Sec}$ which takes 2sec,3sec and 4sec audio data respectively during training. Our final experiments study the effectiveness of using raw waveform methods instead of handcrafted features. We set up an experiment to train the model using MFCC features as inputs instead of the raw waveform to our model. We extract 13 dimensional MFCC (with delta and double-delta) feature for every 25ms using a 10ms frameshift for this experiment. The MFCC based model is referred to as $\textit{ResNet-LSTM-MHA-MFCC}$ while the raw waveform based model is referred to as $\textit{ResNet-LSTM-MHA-Raw}$ in this paper. We use Adam [24] optimizer to train all our models with a learning rate of 0.001 for up to 25 epochs. We use a batch size of 64 during training. We train all our models using Pytorch\footnote{https://pytorch.org/} toolkit.

\section{RESULTS}
\label{sec:illust}
In this section, we describe the evaluation of different models and their performances. We train 2 baseline models $\textit{ResNet}$ and $\textit{ResNet-LSTM}$. The first baseline model $\textit{ResNet}$ consists fo a sequence of 3 Residual blocks made up of 1D convolution kernels. We can think of this model as a ResNet [23] with an average pooling layer replaced by a statistics pooling layer. This model takes 4sec raw audio data and predicts the language label. The $\textit{ResNet}$ model has an F1-score of 88.67\% on the test dataset. The second baseline model $\textit{ResNet-LSTM}$ consists of the same setting as baseline-1, but it has an extra LSTM layer on top of CNN in order to capture long-range temporal information. The performance of this model is 92.21\%, as shown in Table 2. We compare our baseline models with our proposed model $\textit{ResNet-LSTM-MHA-RAW}$, which contains a multi-head attention layer and operates with the raw waveform as input. Table 2 shows that our model gets 3.69\% absolute improvement in F1-score compared to the second baseline model. We also create a model that takes MFCC features as input instead of raw audio refer to as  $\textit{ResNet-LSTM-MHA-MFCC}$. We show that raw waveform based models can get 1.7\% improvement over handcrafted feature-based models.

\begin{table}[!htbp]
  \centering
  \label{tab:tasks}
  \begin{adjustbox}{max width=\textwidth}
    \begin{tabular}{lcc}
      \toprule
      \textbf{System} & \textbf{F1-Score}\\
      \midrule
      $\textit{ResNet}$ (baseline-1) & 89.67\%\\
      $\textit{ResNet-LSTM}$ (baseline-2) & 92.21\%\\
      $\textit{ResNet-LSTM-MHA-MFCC}$ (ours) & 94.22\%\\
      $\textit{ResNet-LSTM-MHA-RAW}$ (ours) & \textbf{95.90}\%\\
      
      \bottomrule
    \end{tabular}
  \end{adjustbox}
  \caption{Comparison of different architectures for language identification. Bold indicates 
  the best performance}
\end{table}

In order to see the effect of the input length during training, we conduct an experiment to train the model using 2Sec, 3Sec, and 4Sec audio data, and we refer to these models as $\textit{ResNet-LSTM-MHA-2Sec}$, $\textit{ResNet-LSTM-MHA-3Sec}$ and $\textit{ResNet-LSTM-MHA-4Sec}$ respectively. The results of these experiments are shown in Table 3. It shows that longer audio data tends to improve the F1-score on the test data due to longer context signals.

\begin{table}[!htbp]
  \centering
  \caption{Comparison of models trained with different segment duration. Bold indicates 
  the best performance}
  \label{tab:tasks}
  \begin{adjustbox}{max width=\textwidth}
    \begin{tabular}{lcc}
      \toprule
      \textbf{System} & \textbf{F1-Score}\\
      \midrule
      $\textit{ResNet-LSTM-MHA-2Sec}$  & 92.64\%\\
      $\textit{ResNet-LSTM-MHA-3Sec}$ & 94.40\%\\
      $\textit{ResNet-LSTM-MHA-4Sec}$ & \textbf{95.90}\%\\
      \bottomrule
    \end{tabular}
  \end{adjustbox}
\end{table}

Finally, we visualize the utterance level embeddings extracted from the projection layer for all the languages. We extract embeddings for 6500 randomly selected test utterances for t-sne visualization. Each embedding has dimension. We reduce the dimension of the embeddings to 2 using the t-sne technique. The t-sne plot of the 2-D embeddings is shown in Figure 2.  It can be clearly seen that the proposed model learns very good language discriminative features at the segment level.

\begin{figure}[t]
  \centering
  \includegraphics[width=\linewidth]{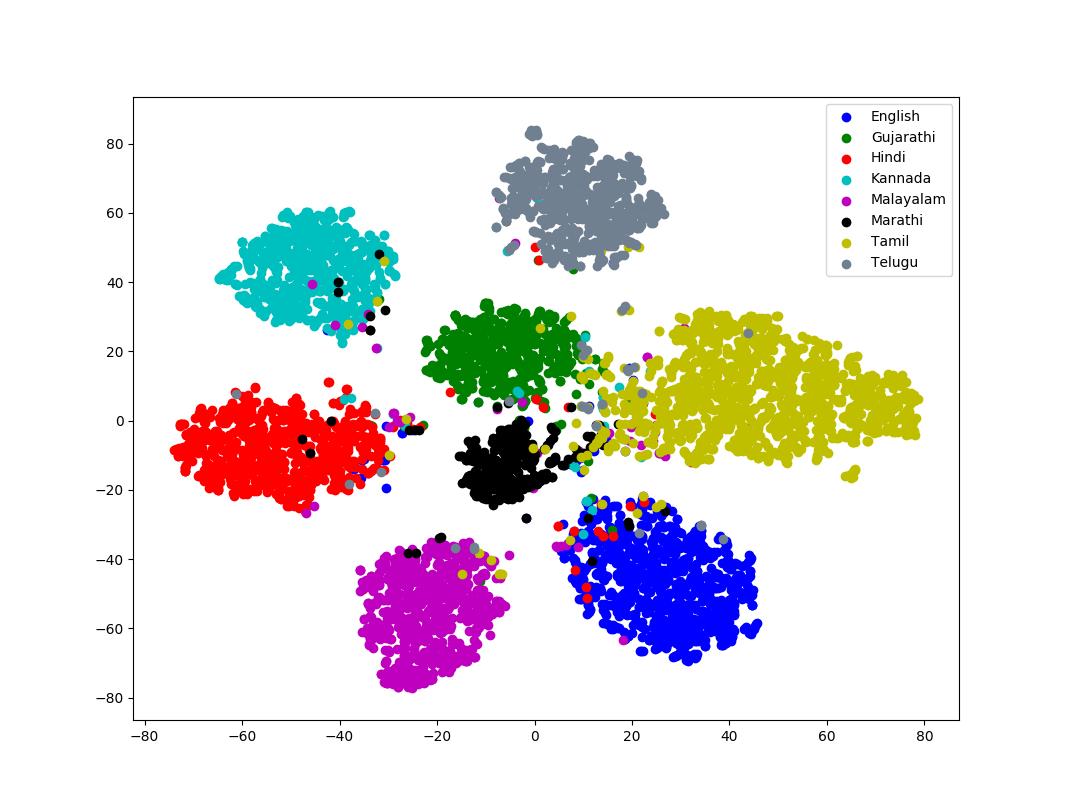}
  \caption{t-sne plot of utterence level embeddings}
  \label{fig:speech_production}
\end{figure}

\section{Conclusion}
\label{sec:foot}
In this work, we propose a new architecture for language identification using multi-head self-attention and 1D convolutional neural networks. We propose to use raw waveform directly as input instead of handcrafted features to learn language discriminative feature using 1D convolution operations. Our model uses multi-head self-attention to learn and select more important features for language identification task.  We finally use a statistics pooling approach to extract utterance level language representation from frame-level features. We collect and curate ~373hrs audio data for 8 Indian languages \textit{Hindi}, \textit{English}, \textit{Kannada},\textit{Tamil}, \textit{Telugu}, \textit{Malayalam}, \textit{Gujarati} and \textit{Marathi}. Our experiments show that multi-head self-attention in combination with raw waveform based 1D convolutional neural network model obtains the best performance on our evaluation dataset. We extract the utterance level embeddings for our evaluation data and visualize the clustering effect using t-sne. The visualization clearly shows that the model learns very good language discriminative features.


\begin{thebibliography}{30}
\newcommand{\quotes}[1]{``#1''}

\bibitem{}
W. Chan, N. Jaitly, Q. V. Le, and O. Vinyals, \quotes{Listen, Attend
and Spell: A Neural Network for Large Vocabulary Conversational
Speech Recognition} in \textit{ICASSP}, 2016

\bibitem{}
S. Kim, T. Hori, and S. Watanabe, \quotes{Joint CTC-attention based endto-end speech recognition using multi-task learning,} in \textit{IEEE International Conference on Acoustics, Speech and Signal Processing (ICASSP)}, 2017, pp. 4835–4839.

\bibitem{}
Ronan Collobert, Christian Puhrsch, and Gabriel Synnaeve,
\quotes{Wav2letter: an end-to-end convnet based speech recognition system,} \textit{CoRR}, vol.abs/1609.03193, 2016.

\bibitem{}
W. Geng, J. Li, S. Zhang, X. Cai, and B. Xu, \quotes{Multilingual tandem
bottleneck feature for language identification,} in \textit{Sixteenth Annual
Conference of the International Speech Communication Association,} 2015.

\bibitem{}
N. Dehak, P. Kenny, R. Dehak, P. Dumouchel, and P. Ouellet,
\quotes{Front-end factor analysis for speaker verification,} \textit{IEEE Transactions on Audio, Speech,and Language Processing} vol. 19, no. 4,pp. 788–798, 2011.

\bibitem{}
Y. Song, B. Jiang, Y. Bao, S. Wei, and L.-R. Dai, \quotes{I-vector representation based on bottleneck features for language identification,}
\textit{Electronics Letters,} vol. 49, no. 24, pp. 1569–1570, 2013

\bibitem{}
I. Lopez-Moreno, J. Gonzalez-Dominguez, O. Plchot, D. Martinez,
J. Gonzalez-Rodriguez, and P. Moreno, \quotes{Automatic language identification using deep neural networks}, in \textit{Acoustics, Speech and Signal Processing (ICASSP), 2014 IEEE International Conference on. IEEE}, 2014, pp. 5337–5341

\bibitem{}
S. Hochreiter and J. Schmidhuber, \quotes{Long short term memory}, \textit{Neural computation}, 1997.

\bibitem{}
J. Gonzalez-Dominguez, I. Lopez-Moreno, H. Sak, J. Gonzalez Rodriguez, and P. J. Moreno, \quotes{Automatic language identification using long short-term memory recurrent neural networks.} in \textit{INTERSPEECH}, 2014, pp. 2155–2159.

\bibitem{}
S. Ganapathy, K. J. Han, S. Thomas, M. K. Omar, M. Van Segbroeck, and S. S. Narayanan, \quotes{Robust language identification using convolutional neural network features.} \textit{ISCA INTERSPEECH,}
2014.

\bibitem{}
Lozano Diez, A. Zazo Candil, R. Gonzalez Dominguez, J.,
Toledano, D.T., and Gonzalez Rodrıguez, J. (2015), \quotes{An end-to-end approach to language identification in short utterances using convolutional neural networks}, \textit{INTERSPEECH 2015}

\bibitem{}
Garcia-Romero, D., McCree, A. \quotes{Stacked Long-Term TDNN for Spoken Language Recognition,} \textit{Proc. Interspeech 2016}, 3226-
3230.

\bibitem{}
D. Snyder, D. Garcia-Romero, G. Sell, D. Povey, and S. Khudanpur,\quotes{X-vectors: Robust dnn embeddings for speaker recognition,} in 2018 \textit{IEEE International Conference on Acoustics, Speech and Signal Processing (ICASSP). IEEE}, 2018

\bibitem{}
D. Snyder, D. Garcia-Romero, A. McCree, G. Sell, D. Povey, and
S. Khudanpur, \quotes{Spoken language recognition using x-vectors,} in
\textit{Odyssey: The Speaker and Language Recognition Workshop}, 2018

\bibitem{}
C Bartz, T Herold, H Yang, and C Meinel ,\quotes{Language identification using deep convolutional recurrent neural networks}, \textit{CoRR},abs/1708.04811, 2017

\bibitem{}
D. Bahdanau, K. Cho, and Y. Bengio. \quotes{Neural machine translation by jointly learning to align and translate}, \textit{arXiv preprint} arXiv:1409.0473, 2014

\bibitem{}
J. Chorowski, D. Bahdanau, K. Cho, and Y. Bengio, \quotes{End-to-end
continuous speech recognition using attention-based recurrent nn:
first results,} \textit{arXiv preprint} arXiv:1412.1602, 2014

\bibitem{}
I. Sutskever, O. Vinyals, and Q. V. Le, \quotes{Sequence to sequence
learning with neural networks,} in \textit{Advances in neural information
processing systems}, 2014, pp. 3104–3112.

\bibitem{}
M. Ravanelli and Y.Bengio, \quotes{Speaker Recognition from raw
waveform with SincNet,} in \textit{Proc. of SLT}, 2018

\bibitem{}
M. Sarma, P. Ghahremani, D. Povey, N. K. Goel, K. K. Sarma, and
N. Dehak, \quotes{Emotion identication from raw speech signals using
dnns}, \textit{Proc. Interspeech} 2018, pp. 3097–3101, 2018.

\bibitem{}
V. Peddinti, D. Povey, and S. Khudanpur, \quotes{A time delay neural
network architecture for efficient modeling of long temporal contexts,} in \textit{Proceedings of INTERSPEECH}, 2015.

\bibitem{}
A. Vaswani, N. Shazeer, N. Parmar, J. Uszkoreit, L. Jones, A. N.
Gomez, Ł. Kaiser, and I. Polosukhin, \quotes{Attention is all you need,}
in \textit{Advances in Neural Information Processing Systems}, 2017, pp.
5998–6008

\bibitem{}
K. He, X. Zhang, S. Ren, and J. Sun. \quotes{Deep residual learning
for image recognition,} \textit{Computer Vision and Pattern Recognition
(CVPR)}, 2016

\bibitem{}
Diederik P. Kingma and Jimmy Ba, \quotes{Adam: A Method for
Stochastic Optimization}, In \textit{Proceedings of the International Conference on Learning Representations (ICLR)}, 2014

\bibitem{}
Okabe, Koji and Koshinaka, Takafumi and Shinoda, Koichi,\quotes{Attentive Statistics Pooling for Deep Speaker Embedding}, In \textit{Interspeech} 2018


\end{thebibliography}
\end{document}